\documentclass[twoside,12pt]{article}
\usepackage{epsfig}

\newcommand{\be}{\begin{equation}}
\newcommand{\ee}{\end{equation}}
\newcommand{\bea}{\begin{eqnarray}}
\newcommand{\eea}{\end{eqnarray}}

\usepackage{amsmath}
\usepackage{amsfonts}
\usepackage{amssymb}
\usepackage{amsxtra}
\usepackage{tikz}
\usetikzlibrary{arrows,snakes,decorations.pathmorphing}
\usepackage{supertabular}

\title{  \vspace{1cm} 
How to present and interpret the Feynman diagrams in this theory 
describing fermion and boson fields in a unique way, in comparison with 
the Feynman diagrams so far presented and interpreted?
}
\author{N.S.\ Manko\v c Bor\v stnik$^{1}$, H.B.\ Nielsen$^{2}$
\\ 
$^1$Department of Physics, University of Ljubljana, 
SI-1000 Ljubljana, Slovenia\\
$^2$Niels Bohr Institute, University of Copenhagen, Blegdansvej 17,
Denmark}  
\date{}
\begin{document}
\maketitle
\begin{abstract}
Although the internal spaces describing spins and charges of fermions'
and bosons' second-quantised fields have such different properties, yet we can all describe them equivalently with the ``basis vectors'' which are a
superposition of odd (for fermions) and even (for bosons) products of
$\gamma^{a}$'s. In an even-dimensional internal space, as it is
$d=(13+1)$, odd ``basis vectors'' appear in $2^{\frac{d}{2}-1}$
families with $2^{\frac{d}{2}-1}$ members each, and have their
Hermitian conjugate partners in a separate group, while even
``basis vectors'' appear in two orthogonal groups. Algebraic
multiplication of boson and fermion ``basis vectors'' determine the
interactions between fermions and bosons, and among bosons
themselves, and correspondingly also their action. Tensor products of
the ``basis vectors'' and basis in ordinary space-time determine states
for fermions and bosons, if bosons obtain in addition the space index
$\alpha$. %
We study properties of massless fermions and bosons with 
the internal spaces determined by the ``basis vectors'' while assuming 
that fermions and bosons are active only in $d=(3+1)$ of the ordinary 
space-time.
We discuss the Feynman diagrams in this theory, describing internal 
spaces of fermion and boson fields with odd and even ``basis vectors'', 
respectively, in comparison with the Feynman diagrams of the theories 
so far presented and interpreted.
\end{abstract}

\section{Introduction}
\label{introduction}
Authors studied (together, and with the collaborators) in a series of 
papers the properties of the second quantized fermion and boson fields~%
\cite{norma93,norma95,pikanorma2005,nh00,nh02,nd2017,nh2021RPPNP,
n2024IARD,n2023NPB,n2024NPB,n2023MDPI,gmdn2008,gn2009,
n2014matterantimatter,JMP2013,gn2014,nm2015}, trying to understand 
what are the elementary laws of nature for massless fermion and boson 
fields, and whether all the second quantized fields, fermions' and bosons', 
can be described in an unique and simple way.


Accepting the idea of the papers~\cite{n2023NPB,n2024NPB,n2023MDPI,n2024IARD}
that internal spaces of fermions and bosons are described by ``basis
vectors'' which are the superposition of odd (for fermions) and even (for
bosons) products of the operators $\gamma^{a}$'s, the authors continue
to find out whether and to what extent ``nature manifests'' the
proposed idea.

As presented in one contribution of this proceedings~(\cite{n2025Bled}, the 
talk of one of the two authors), the idea
that the internal spaces of fermion and boson fields are described by the
odd and even ``basis vectors'' which are products of nilpotents and
projectors, all of which are the eigenvectors of the Cartan subalgebra
members of the Lorentz algebra in the internal space of the fermion and
boson fields, enabling to explain the second quantisation postulates of
Dirac determines uniquely in even-dimensional spaces with $d=2(2n+1)$
also the action for interacting massless fermion, antifermion and boson
fields.

In Sect.~\ref{feynman}, we present and comment on the Feynman diagrams
for interacting fermions and bosons when describing internal spaces with
our proposal, paying attention to massless fermions and bosons,
and for fermions and bosons in ordinary theories.

In Sect.~\ref{openproblems} we comment on the results of our presentation.

In the introduction, we overview:\\
{\bf a. \,\,\,} The construction of the odd and even ``basis vectors''
describing the internal spaces of fermions and bosons.\\
{\bf b. \,\,\,} The algebraic products among fermion and boson ``basis
vectors'' which determine the action for both fields and interactions among
them. \\
{\bf c. \,\,\,} The tensor products of ``basis vectors'' and the basis in
ordinary space-time, determining the massless anticommuting fermion and
commuting boson second quantised fields, in which bosons gain the
vector index $\alpha$, while fermions and bosons are active (have non-zero
momentum) only in $d=(3+1)$, while the internal spaces have $d=(13+1)$.
Bosons can gain a vector index $\mu=(0,1,2,3)$, representing gravitons of
spin $\pm 2$, vectors of spin $\pm 1$ (photons, weak bosons, gluons)
or a scalar index $\sigma=(5,6,...,13)$, representing scalars (Higgs and
others).

In Sect.~\ref{feynman}, we present and comment on the Feynman diagrams
for interacting fermions and bosons when describing internal spaces with
our proposal, paying attention to massless fermions and bosons,
and for fermions and bosons in ordinary theories.

In Sect.~\ref{openproblems} we comment on the results of our presentation.
 


%
\subsection{States of the second quantized fermion and boson fields}
\label{fermionbosonstates}
This Subsect.~\ref{fermionbosonstates} is a short overview of several 
similar sections, presented in Refs.~\cite{n2023NPB,n2024NPB,%
nh2021RPPNP,n2025Bled}, the last one, Ref.~\cite{n2025Bled}, appears 
in this Proceedings.

In this contribution, all the  second quantised fermion and boson states 
are assumed to be massless. They are constructed as tensor products
of ``basis vectors'', which determine the anti-commutation properties 
of fermions and commutation properties of bosons, also in the tensor 
product with basis in ordinary space-time. We present the ``basis 
vectors'' as products of nilpotents and projectors, so that they are 
eigenstates of all the Cartan subalgebra members of the Lorentz algebra
in the $d=(13+1)$-dimensional internal space, while the space-time 
has only $d=(3+1)$.


The Grassmann algebra offers two kinds of operators $\gamma^a$'s%
~\cite{n2023NPB,n2024NPB,nh2021RPPNP,n2025Bled}, we call them
$\gamma^a$'s and $\tilde{\gamma}^a$'s with the properties
\begin{eqnarray}                             
\label{gammatildeantiher}
\{\gamma^{a}, \gamma^{b}\}_{+}&=&2 \eta^{a b}=
\{\tilde{\gamma}^{a},
\tilde{\gamma}^{b}\}_{+}\,, \nonumber\\
\{\gamma^{a}, \tilde{\gamma}^{b}\}_{+}&=&0\,,\quad
(a,b)=(0,1,2,3,5,\cdots,d)\,, \nonumber\\
(\gamma^{a})^{\dagger} &=& \eta^{aa}\, \gamma^{a}\, ,
\quad (\tilde{\gamma}^{a})^{\dagger} = \eta^{a a}\,
\tilde{\gamma}^{a}\,.
\end{eqnarray}
We use $\gamma^a$'s, to generate the ``basis vectors'' describing 
internal spaces of fermions and bosons, arranging them to be products 
of nilpotents and projectors
\begin{small}
\begin{eqnarray}
\label{nilproj}
\stackrel{ab}{(k)}:&=&\frac{1}{2}(\gamma^a +
\frac{\eta^{aa}}{ik} \gamma^b)\,, \;\;\; (\stackrel{ab}{(k)})^2=0\, ,
\nonumber \\
\stackrel{ab}{[k]}:&=&
\frac{1}{2}(1+ \frac{i}{k} \gamma^a \gamma^b)\,, \;\;\;
(\stackrel{ab}{[k]})^2=
\stackrel{ab}{[k]}\,.
\end{eqnarray}
\end{small}
Nilpotents are a superposition of an odd number of $\gamma^a$'s, projectors
of an even number of $\gamma^a$'s, both are chosen to be the eigenstate
of one of the (chosen) Cartan subalgebra members of the Lorentz algebra of
$S^{ab}= \frac{i}{4} \{\gamma^a, \gamma^b\}_+$, and
$\tilde{S}^{ab}=  \frac{i}{4} \{\tilde{\gamma}^a,
\tilde{\gamma}^b\}_+$ in the internal space of fermions and bosons.
\begin{small}
\begin{eqnarray}
&&S^{03}, S^{12}, S^{56}, \cdots, S^{d-1 \;d}\,, \nonumber\\
&&\tilde{S}^{03}, \tilde{S}^{12}, \tilde{S}^{56}, \cdots,  \tilde{S}^{d-1\; d}\,,
\nonumber\\
&&{\cal {\bf S}}^{ab} = S^{ab} +\tilde{S}^{ab}\,.
\label{cartangrasscliff}
\end{eqnarray}
\end{small}
%
%
\begin{small}
\begin{eqnarray}
\label{calsab}
S^{ab} \,\stackrel{ab}{(k)} = \frac{k}{2} \,\stackrel{ab}{(k)}\,,\quad && \quad
\tilde{S}^{ab}\,\stackrel{ab}{(k)} = \frac{k}{2} \,\stackrel{ab}{(k)}\,,\nonumber\\
S^{ab}\,\stackrel{ab}{[k]} = \frac{k}{2} \,\stackrel{ab}{[k]}\,,\quad && \quad
\tilde{S}^{ab} \,\stackrel{ab}{[k]} = - \frac{k}{2} \,\,\stackrel{ab}{[k]}\,,
\end{eqnarray}
\end{small}
with $k^2=\eta^{aa} \eta^{bb}$.

In even-dimensional spaces, the states in internal spaces are defined by
the ``basis vectors'' which are products of  $\frac{d}{2}$ nilpotents and
projectors, and are the eigenstates of all the Cartan subalgebra
members.\\

\vspace{2mm}

{\bf a.\,\,\,}
``Basis vectors'' including an odd number of nilpotents (at least one, the 
rest are projectors) anti-commute, since the odd products of $\gamma^a$'s 
anti-commute. The odd ``basis vectors'' are used to describe 
fermions. The odd ``basis vectors'' appear in internal  spaces with $d=2(2n+1)$ 
in $2^{\frac{d}{2}-1}$ irreducible representations, called families, with the 
quantum numbers determined by $\frac{d}{2}$ members of 
Eq.~(\ref{cartangrasscliff}). Each family has $2^{\frac{d}{2}-1}$ members.
$S^{ab}$ transform family members within each family. $\tilde{S}^{ab}$ 
transform a family member of one family to the same family member of 
the rest of family. The Hermitian conjugated partners of the odd ``basis 
vectors'' have $2^{\frac{d}{2}-1}\times 2^{\frac{d}{2}-1}$ members and 
appear in a different group. The odd ``basis vectors'' and their Hermitian 
conjugated partners have together $2^{d -1}$ members. \\

We call the odd ``basis vectors'' $\hat{b}^{m \dagger}_{f}$, and their 
Hermitian conjugated partners $\hat{b}^{m }_{f}=$
$(\hat{b}^{m \dagger}_{f})^{\dagger}$. $m$ denotes the membership
and $f$ the family quantum number of the odd ``basis vectors''. \\

The algebraic product, $*_{A}$, of any two members of the odd ``basis 
vectors'' are equal to zero. And any two members of their Hermitian 
conjugated partners have the algebraic product, $*_{A}$, equal to zero.

\begin{eqnarray}
\hat{b}^{m \dagger}_f *_{A} \hat{b}^{m `\dagger }_{f `}&=& 0\,,
\quad \hat{b}^{m}_f *_{A} \hat{b}^{m `}_{f `}= 0\,, \quad \forall m,m',f,f `\,.
\label{orthogonalodd}
\end{eqnarray}
%

Choosing the vacuum state equal to
\begin{eqnarray}
\label{vaccliffodd}
|\psi_{oc}>= \sum_{f=1}^{2^{\frac{d}{2}-1}}\,\hat{b}^{m}_{f}{}_{*_A}
\hat{b}^{m \dagger}_{f} \,|\,1\,>\,,
\end{eqnarray}
for one of the members $m$, anyone of the odd irreducible representations $f$, 
it follows that the odd ``basis vectors'' obey the relations
\begin{small}
\begin{eqnarray}
\label{almostDirac}
\hat{b}^{m}_{f} {}_{*_{A}}|\psi_{oc}>&=& 0.\, |\psi_{oc}>\,,\nonumber\\
\hat{b}^{m \dagger}_{f}{}_{*_{A}}|\psi_{oc}>&=&  |\psi^m_{f}>\,,\nonumber\\
\{\hat{b}^{m}_{f}, \hat{b}^{m'}_{f `}\}_{*_{A}+}|\psi_{oc}>&=&
 0.\,|\psi_{oc}>\,, \nonumber\\
\{\hat{b}^{m \dagger}_{f}, \hat{b}^{m' \dagger}_{f  `}\}_{*_{A}+}|\psi_{oc}>
&=& 0. \,|\psi_{oc}>\,,\nonumber\\
\{\hat{b}^{m}_{f}, \hat{b}^{m' \dagger}_{f `}\}_{*_{A}+}|\psi_{oc}>
&=& \delta^{m m'} \,\delta_{f f `}|\psi_{oc}>\,,
\end{eqnarray}
\end{small}\\
as postulated by Dirac for the second quantised fermion fields. In 
Eq.~(\ref{almostDirac}) odd ``basis vectors'' anti-commute, since $\gamma^a$'s 
obey Eq.~(\ref{gammatildeantiher}).

Being eigenstates of operators $S^{ab}$ and $\tilde{S}^{ab}$, when $(a,b)$
belong to Eq.~(\ref{cartangrasscliff}), nilpotents and projectors carry both 
quantum numbers $S^{ab}$ and $\tilde{S}^{ab}$, Eq.~(\ref{cartangrasscliff}).

$S^{ab}$ transform the odd ``basis vectors'' of family $f$ to all the members 
of the same family, $\tilde{S}^{ab}$ transform a particular family member to 
the same family member of all the families.\\

\vspace{2mm}

{\bf b.\,\,\,} The even ``basis vectors'' commute, since the even products 
of $\gamma^a$'s commute, Eq.~(\ref{gammatildeantiher}).  
In internal  spaces with $d=2(2n+1)$, the even ``basis vectors'' appear 
in two orthogonal groups. We name them  
${}^{I}\hat{\cal A}^{m \dagger}_{f}$ and 
${}^{II}\hat{\cal A}^{m \dagger}_{f}$.
\begin{eqnarray}
\label{AIAIIorth}
{}^{I}{\hat{\cal A}}^{m \dagger}_{f} *_A {}^{II}{\hat{\cal A}}^{m \dagger}_{f}
&=&0={}^{II}{\hat{\cal A}}^{m \dagger}._{f} *_A
{}^{I}{\hat{\cal A}}^{m \dagger}_{f}\,.
\end{eqnarray}
Each group has $2^{\frac{d}{2}-1}\times 2^{\frac{d}{2}-1}$ members 
with the Hermitian conjugate partners within the group.

The even ``basis vectors'' have the eigenvalues of the Cartan subalgebra 
members, Eq.~(\ref{cartangrasscliff}), equal to ${\cal S}^{ab}=(S^{ab}+ 
\tilde{S}^{ab})$, their eigenvalues are $\pm i$ or $\pm 1$ or zero. According to
Eq.~(\ref{calsab}), the eigenvalues of ${\cal S}^{ab}$ are for projectors equal
zero; ${\cal S}^{ab}(= S^{ab}+ \tilde{S}^{ab})\,\stackrel{ab}{[\pm]}=0$.

The algebraic products, $ \,*_A\,$, of two members of each of these two groups
 have the property
\begin{eqnarray}
\label{ruleAAI}
{}^{i}{\hat{\cal A}}^{m \dagger}_{f} \,*_A\, {}^{i}{\hat{\cal A}}^{m' \dagger}_{f `}
\rightarrow \left \{ \begin{array} {r}
{}^{i}{\hat{\cal A}}^{m \dagger}_{f `}\,, i=(I,II) \\
{\rm or \,zero}\,.
\end{array} \right.
\end{eqnarray}
$i$ is either $I$ or $II$.
For a chosen ($m, f, f `$), there is (out of $2^{\frac{d}{2}-1}$) only one $m'$  
giving a non-zero contribution.

We further find
\begin{eqnarray}
\label{calIAb1234gen}
{}^{I}{\hat{\cal A}}^{m \dagger}_{f } \,*_A \, \hat{b}^{m' \dagger }_{f `}
\rightarrow \left \{ \begin{array} {r} \hat{b }^{m \dagger}_{f `}\,, \\
{\rm or \,zero}\,.
\end{array} \right.
\end{eqnarray}
Eq.~(\ref{calIAb1234gen}) demonstrates that
${}^{I}{\hat{\cal A}}^{m \dagger}_{f}$,
applying on $\hat{b}^{m' \dagger }_{f `} $, transforms the odd
``basis vector'' into another odd ``basis vector'' of the same family,
transferring to the odd ``basis vector'' integer spins or gives zero.

For the second group of boson fields, ${}^{II}{\hat{\cal A}}^{m \dagger}_{f }$, 
it follows 
\begin{eqnarray}
\label{calbIIA1234gen}
\hat{b}^{m \dagger }_{f } *_{A} {}^{II}{\hat{\cal A}}^{m' \dagger}_{f `} \,
\rightarrow \left \{ \begin{array} {r} \hat{b }^{m \dagger}_{f ``}\,, \\
{\rm or \,zero}\,.
\end{array} \right.
\end{eqnarray}
The application of the odd ``basis vector'' 
$\hat{b}^{m \dagger }_{f }$ on 
$ {}^{II}{\hat{\cal A}}^{m' \dagger}_{f `}$ leads to another 
odd ``basis vector'' $\hat{b }^{m \dagger}_{f ``}$ belonging 
to the same family member $m$ of a different family $f ``$.

The rest of possibilities give zero.

Knowing the odd ``basic vectors'', we can generate all the even ``basic 
vectors'' 
\begin{eqnarray}
\label{AIbbdagger}
{}^{I}{\hat{\cal A}}^{m \dagger}_{f}&=&\hat{b}^{m' \dagger}_{f `} *_A 
(\hat{b}^{m'' \dagger}_{f `})^{\dagger}\,,
\end{eqnarray}
\begin{eqnarray}
\label{AIIbdaggerb}
 {}^{II}{\hat{\cal A}}^{m \dagger}_{f}&=&
(\hat{b}^{m' \dagger}_{f `})^{\dagger} *_A 
\hat{b}^{m' \dagger}_{f `'}\,. 
\end{eqnarray}\\

\vspace{3mm}

{\bf c. \,\,\,} To define the fermion and boson second quantized fields 
we must write the tensor product, $\, *_{T}\,$ of the ``basis vectors''
in internal space with $d=(13+1)$ and the ordinary space-time in the 
case fermions and bosons have non-zero momentum only in
$d=(3+1)$. For boson fields, we need to postulate the space index 
$\alpha$, which is for vectors (representing gravitons, photons, weak 
bosons, gluons) equal to $\mu=(0,1,2,3)$  and for scalars equal to 
$\sigma \ge 5$.\\

Let us start with basis in ordinary space-time, following 
Refs.~\cite{nh2021RPPNP,n2025Bled,n2023NPB,n2024NPB}.
\begin{eqnarray}
\label{creatorp}
|\vec{p}>&=& \hat{b}^{\dagger}_{\vec{p}} \,|\,0_{p}\,>\,,\quad
<\vec{p}\,| = <\,0_{p}\,|\,\hat{b}_{\vec{p}}\,, \nonumber\\
<\vec{p}\,|\,\vec{p}'>&=&\delta(\vec{p}-\vec{p}')=
<\,0_{p}\,|\hat{b}_{\vec{p}}\; \hat{b}^{\dagger}_{\vec{p}'} |\,0_{p}\,>\,,
\nonumber\\
<\,0_{p}\,| \hat{b}_{\vec{p'}}\, \hat{b}^{\dagger}_{\vec{p}}\,|\,0_{p}\, > 
&=&\delta(\vec{p'}-\vec{p})\,,
\end{eqnarray}
with $<\,0_{p}\, |\,0_{p}\,>=1$. The operator 
$\hat{b}^{\dagger}_{\vec{p}}$ pushes a  single particle state 
with zero momentum by an amount $\vec{p}$.\\

The creation operator for a free massless fermion field of the energy
$p^0 =|\vec{p}|$, belonging to the family $f$ and to a superposition of
family members $m$ applying on the vacuum state
($|\psi_{oc}>\,*_{T}\, |0_{\vec{p}}>$) 
can be written as
\begin{small}
\begin{eqnarray}
\label{wholespacefermions}
{\bf \hat{b}}^{m \dagger}_{f} (\vec{p}) \,&=& 
\,\hat{b}^{\dagger}_{\vec{p}}\,*_{T}\,
\hat{b}^{m \dagger}_{f} \,.
\end{eqnarray}
\end{small}\\

The creation operator for a free massless boson field of the energy
$p^0 =|\vec{p}|$,  with the ``basis vectors'' belonging to one of the 
two groups, ${}^{i}{\hat{\cal A}}^{m \dagger}_{f }, i=(I,II)$,
applying on the vacuum state, $|\,1\,>\,*_{T}\, |0_{\vec{p}}>$, 
carrying the space index $\alpha$,  we have
\begin{eqnarray}
\label{wholespacebosons}
{\bf {}^{i}{\hat{\cal A}}^{m \dagger}_{f a}} (\vec{p}) \,&=&
{}^{i}{\cal C}^{ m}{}_{f a} (\vec{p})\,*_{T}\,
{}^{i}{\hat{\cal A}}^{m \dagger}_{f} \, \,, i=(I,II)\,, (f,m) 
\end{eqnarray}
with ${}^{i}{\cal C}^{ m}{}_{f a} (\vec{p})=
{}^{i}{\cal C}^{ m}{}_{f a}\,\hat{b}^{\dagger}_{\vec{p}}$\,,
and $ (f,m) $ are fixed values, the same on both sides.\\


Le us add that the Lorentz rotations work on both spaces  only in 
$d=(3+1)$.\\

\vspace{3mm}

{\bf d. \,\,\,}  Knowing the application among fermion and boson 
``basis vectors'', from Eq.~(\ref{AIAIIorth}) to Eq.~(\ref{AIIbdaggerb}),
we can write down the action
\begin{eqnarray}
\label{action}
{\cal A}\,  &=& \int \; d^4x \;\frac{1}{2}\, (\bar{\psi} \, \gamma^a p_{0a} \psi)
+ h.c. +
\nonumber\\
& & \int \; d^4x \;\sum_{i=(I,II)}\,
{}^{i}{\hat F}^{m\,f}_{ab}\;\, {}^{i}{\hat F}^{m f ab}\,,
\nonumber\\
p_{0a}  &=& p_{a}  -
\sum_{m f}   {\bf {}^{I}{\hat{\cal A}}^{m \dagger}_{f a}}(x) -
\sum_{m f}  {\bf {}^{II}{\hat{\cal A}}^{m \dagger}_{f a}}(x) \, \,,
\nonumber\\
{}^{i}{\hat F}^{m\,f}_{ab} &=&  \partial_{a}\,
{\bf {}^{i}{\hat{\cal A}}^{m \dagger}_{f b}}(x) - \partial_{b}\,
{\bf {}^{i}{\hat{\cal A}}^{m \dagger}_{f a}}(x) + \varepsilon
{\bf f}^{m f m'' f \,'' m' f `}\,{\bf {}^{i}{\hat{\cal A}}^{m'' \dagger}_{f \,'' a}}(x)
\,{\bf {}^{i}{\hat{\cal A}}^{m' \dagger}_{f `b}}(x) \,,\nonumber\\
&& i=(I,II) \,.
\end{eqnarray}
Vector boson fields, ${}^{i}{\hat{\cal A}}^{m \dagger}_{f a}$ (and 
in ${}^{i}{\hat F}^{m\,f}_{ab}$), must have index $(a,b)$ equal to
$(n,p)=(0,1,2,3)$;  ${}^{i}{\hat{\cal A}}^{m \dagger}_{f n}$ (and 
in ${}^{i}{\hat F}^{m\,f}_{np}$), $i=(I,II)$.\\




%
\section{Feynman diagrams in our way and in the way with ordinary theories}
\label{feynman}
%
This section studies the Feynman diagrams in the case when the ``basis
vectors" describe the internal spaces of fermion and boson fields; the
``basis vectors" of fermions have an odd number of nilpotents, and
those of bosons have an even number of nilpotents, with the rest being
projectors. We compare these Feynman diagrams with those in which 
the internal spaces of fermions and bosons are described by matrices, 
while the fermion families must be postulated, as is the case in most 
theories.

Let it be repeated: We study the scattering of fermion in boson fields, 
which are tensor products of the ``basis vectors'' and basis in ordinary
space-time. ``Basis vectors'' determine spins and charges of fermions
and bosons, families of fermion fields and two kinds of boson fields,
as well as anti-commutativity and commutativity of fields.

In $d=2(2n+1)$, each family of ``basis vectors'' of fermion fields
includes fermions and anti-fermions: $d=(13+1)$ includes quarks and
leptons and anti-quarks and anti-leptons. Quarks have identical
$d=(7+1)$ part  of $d=(13+1)$ as leptons; anti-quarks have identical
$d=(7+1)$ part  of $d=(13+1)$ as anti-leptons.
Quarks are distinguished from leptons and anti-quarks from anti-leptons 
only in the $SO(6)$ part of $SO(13,1)$.

``Basis vectors'' in $d=4n$ include fermions and do not include
anti-fermions; there are no anti-fermions in $d=4n$~\footnote{
Let us look at one family of the fermion ``basis vectors'' in $d=(7+1)$, 
to notice that we do not have members who could represent 
antiparticles with opposite charge and opposite handedness. On the 
left-hand side, the ``basis vectors'' are presented, on the right-hand 
side, their Hermitian conjugate partners. In the case of $d=(7+1)$ 
and when taking care of only the internal spaces of fermions and 
bosons, the discrete symmetry operator 
$\mathbb{C}_{{ \cal N}}{\cal P}^{(d-1)}_{{\cal N}}$, Eq.~(24) 
in~\cite{n2024NPB}, simplifies to $\gamma^0 \gamma^5 
\gamma^7. $  Having odd numbers of operators $\gamma^a$'s, 
it would transform a fermion into a boson. We easily notice that 
there are no pairs, which would have opposite handedness and 
opposite charges.
\begin{small}
\begin{eqnarray}
\label{allcartaneigenvec4n}
&& \qquad \qquad \qquad \qquad d=4n\, ,\nonumber\\
&& \hat{b}^{1 \dagger}_{1}=\stackrel{03}{(+i)}\stackrel{12}{[+]}
\stackrel{5 6}{[+]} \stackrel{7 8}{[+]}\,,\qquad  
\hat{b}^{1}_{1}=\stackrel{03}{(-i)}\stackrel{12}{[+]}
\stackrel{5 6}{[+]} \stackrel{7 8}{[+]}\,
\nonumber\\
&&\hat{b}^{2 \dagger}_{1} = \stackrel{03}{[-i]} \stackrel{12}{(-)}
\stackrel{56}{[+]}\stackrel{78}{[+]}\,,\qquad 
\hat{b}^{2 }_{1} = \stackrel{03}{[-i]} \stackrel{12}{(+)}
\stackrel{56}{[+]}  \stackrel{78}{[+]}\,,
\nonumber\\ 
&&\hat{b}^{3 \dagger}_{1} = \stackrel{03}{(+i)}\stackrel{12}{[+]}
\stackrel{56}{(-)} \stackrel{78}{(-)}\,, \quad\,\,\,
\hat{b}^{3}_{1} =  \stackrel{03}{(-i)}\stackrel{12}{[+]}
\stackrel{56}{(+)} \stackrel{78}{(+)}\,\,,
\nonumber\\
&&\hat{b}^{4 \dagger}_{1} = \stackrel{03}{[-i]} \stackrel{12}{(-)}
\stackrel{56}{(-)} \stackrel{78}{(-)}\,, \quad\,\,\,
\hat{b}^{4}_{1} =   \stackrel{03}{[-i]} \stackrel{12}{(+)}
\stackrel{56}{(+)} \stackrel{78}{(+)}\,\,,
\nonumber\\
&& \hat{b}^{5 \dagger}_{1}=\stackrel{03}{[-i]}\stackrel{12}{[+]}
\stackrel{5 6}{(-)} \stackrel{7 8}{[+]}\,,\qquad  
\hat{b}^{5}_{1}=\stackrel{03}{[-i]}\stackrel{12}{[+]}
\stackrel{5 6}{(+)} \stackrel{7 8}{[+]}\,
\nonumber\\
&&\hat{b}^{6 \dagger}_{1} = \stackrel{03}{(+i)} \stackrel{12}{(-)}
\stackrel{56}{(-)}\stackrel{78}{[+]}\,,\quad \,\,\,
\hat{b}^{6 }_{1} = \stackrel{03}{(-i)} \stackrel{12}{(+)}
\stackrel{56}{(+)}  \stackrel{78}{[+]}\,,
\nonumber\\ 
&&\hat{b}^{7 \dagger}_{1} = \stackrel{03}{[-i]}\stackrel{12}{[+]}
\stackrel{5 6}{[+]} \stackrel{7 8}{(-)}\,, \quad\,\,\,\,\,
\hat{b}^{7}_{1} =  \stackrel{03}{[-i]}\stackrel{12}{[+]}
\stackrel{5 6}{[+]}  \stackrel{78}{(+)}\,\,,
\nonumber\\
&&\hat{b}^{8 \dagger}_{1} = \stackrel{03}{(+i)} \stackrel{12}{(-)}
\stackrel{5 6}{[+]} \stackrel{7 8}{(-)}\,, \quad\,\,\,
\hat{b}^{8}_{1} =   \stackrel{03}{(-i)} \stackrel{12}{(+)}
\stackrel{56}{[+]} \stackrel{78}{(+)}\,\,,
\end{eqnarray}
\end{small}
}. 
To have fermions and anti-fermions, the internal space must be
$d=2(2n+1)$.

Since we assume that fermions and bosons have non-zero momenta only
in $d=(3+1)$ of ordinary space-time, the Lorentz rotations,
$M^{ab}= L^{ab} +S^{ab} +\tilde{S}^{ab}$, connecting both spaces
are possible only in $d=(3+1)$. For $d\ge 5$ the Lorentz rotations
concern only $S^{ab}$ and $\tilde{S}^{ab}$, that is only the internal
space.\\

Let us also point out that since each family in this presentation of the
internal spaces of fermions and bosons includes fermions and anti-fermions,
no negative energy Dirac sea of fermions is needed. The vacuum state is
only the quantum vacuum. Correspondingly, our Feynman diagrams can
differ from the usual ones with the Dirac sea whenever in the diagram 
both the fermion and the anti-fermion appear.~\footnote{
We should also not forget that our second quantised fields, when they
have an odd number of nilpotents, anti-commute; when they have an 
even number of nilpotents, they commute: They are second quantised 
fields needing no postulates.
}
Eq.~(\ref{orthogonalodd}) reminds us that all fermion ``basis vectors''
are orthogonal, and also  their Hermitian conjugate partners are among
themselves orthogonal.

%
\subsection{ ``Basis vectors'' in $d=(5+1)$ and in $d=(13+1)$}
\label{5+1and13+1}

Let us present fermion and boson ``basis vectors'' for some cases,  
$d=(5+1)$ and $d=(13+1)$, to understand better the difference 
between the Feynman diagrams in our case and in most of theories. 

In Table~\ref{Table Clifffourplet.} all odd ``basis vectors'' and their
Hermitian conjugated partners, and all even ``basis vectors'' of two 
kinds are presented. Let us check their properties with respect to
Eqs.~(\ref{orthogonalodd} - \ref{AIIbdaggerb}) to easier follow the
discussions on Feynman diagrams.\\

In Eq.~(\ref{calsab}) we read that either the nilpotents or projectors
carry both quantum numbers $S^{ab}$ and $S^{ab}$. While 
for fermions the first, $S^{ab}$, determines the family member 
quantum number (presented in Table~\ref{Table Clifffourplet.} for 
$\hat{b}^{m \dagger }_{f}$ in the last three columns), and 
$\tilde{S}^{ab}$ the family quantum number (presented in 
Table~\ref{Table Clifffourplet.} for $\hat{b}^{m \dagger }_{f}$ above 
each family), are for bosons the quantum numbers, expressed 
as  ${\cal S}^{ab}= (S^{ab}+\tilde{S}^{ab})$, for nilpotents of integer 
values and for projectors zero.\\

Let us check that the boson ``basis vector'' 
${}^I\hat{{\cal A}}^{4 \dagger}_{1}
(\equiv \stackrel{03}{(+i)}\stackrel{12}{(+)}
\stackrel{56}{[+]})$ is expressible by $\hat{b}^{1 \dagger }_{1} 
(\equiv \stackrel{03}{(+i)}\stackrel{12}{[+]}\stackrel{56}{[+]})
\,*_{A}\, (\hat{b}^{2 \dagger }_{1} )^{\dagger} (\equiv 
\stackrel{03}{[-i]}\stackrel{12}{(+)}
\stackrel{56}{[+]})$.  One can check this by recognizing  that,
$\stackrel{03}{(+i)}\,*_{A}\,\stackrel{03}{[-i]}=\stackrel{03}{(+i)}$, 
$\stackrel{12}{[+]}\,*_{A}\,\stackrel{12}{(+)}=\stackrel{12}{(+)}$
and $\stackrel{56}{[+]}\,*_{A}\,\stackrel{56}{[+]}=\stackrel{56}{[+]}$, 
which can be calculated using Eq.~(\ref{nilproj}), or read in 
Eq.~(\ref{graficcliff0}) of the footnote~\footnote{
\begin{small}
\begin{eqnarray} 
\stackrel{ab}{(k)}\stackrel{ab}{(-k)}& =& \eta^{aa} \stackrel{ab}{[k]}\,,\quad 
\stackrel{ab}{(-k)}\stackrel{ab}{(k)} = \eta^{aa} \stackrel{ab}{[-k]}\,,\quad
\stackrel{ab}{(k)}\stackrel{ab}{[k]} =0\,,\quad 
\stackrel{ab}{(k)}\stackrel{ab}{[-k]} =
 \stackrel{ab}{(k)}\,,\quad 
 \nonumber\\
 \stackrel{ab}{(-k)}\stackrel{ab}{[k]} &=& \stackrel{ab}{(-k)}\,,\quad 
\stackrel{ab}{[k]}\stackrel{ab}{(k)}= \stackrel{ab}{(k)}\,,
\quad 
 \stackrel{ab}{[k]}\stackrel{ab}{(-k)} =0\,,\quad 
 \stackrel{ab}{[k]}\stackrel{ab}{[-k]} =0\,.\quad 
\label{graficcliff0}
 \end{eqnarray}
\end{small}
}.
Using this footnote one easily finds that all odd ``basis vectors''
are orthogonal, as well are orthogonal among themselves all Hermitian 
conjugated partners.\\

If we call $\hat{b}^{1 \dagger }_{1} 
(\equiv \stackrel{03}{(+i)}\stackrel{12}{[+]}\stackrel{56}{[+]})$ the 
fermion with the spin $\uparrow$ having the charge $\frac{1}{2}$ 
($S^{56} \stackrel{56}{[+]}= \frac{1}{2} \stackrel{56}{[+]}$) and 
the right handedness, then we can call 
$\hat{b}^{3 \dagger }_{1} (\equiv \stackrel{03}{[-i]}\stackrel{12}{[+]}
\stackrel{56}{(-)})$ its anti-fermion with the spin $\uparrow$ having the 
charge $-\frac{1}{2}$  and the left handedness. \\

Table~\ref{Table Clifffourplet.}, made for $d=(5+1)$, contains four
families with four odd ``basis vectors'' for fermions. Each family contains
two fermions with the positive charge, $S^{56}=\frac{1}{2}$, one with
the spin up, $\uparrow$, and the other with spin down, $\downarrow$;
and two anti-fermions, again one with the spin up, $\uparrow$, and one
with the spin down, $\downarrow$. The tensor product with the basis
in ordinary space, Eq.~(\ref{wholespacefermions}), represent fermions
and anti-fermions - a kind of electrons and positrons, in this model.

Moreover, we have $16$ corresponding Hermitian conjugate partners.

From these $16$ odd ``basis vectors'', $\hat{b}^{m \dagger }_{f}$, 
and their  $16$ Hermitian conjugated partners, $\hat{b}^{m}_{f}$, we 
construct two groups of $16$ even ``basis vectors'', representing the
internal spaces of bosons, presented in Table~\ref{Table Clifffourplet.}
as ${}^{I}\hat{\cal A}^{m \dagger}_{f}$ and 
${}^{II}\hat{\cal A}^{m \dagger}_{f}$. The tensor products of even 
``basis vectors'' with the basis in the ordinary space-time, and with 
the space index $\alpha =\mu \le 3$ or $\alpha =\sigma \ge 5$,  
Eq.~(\ref{wholespacefermions}), 
represent two kinds of boson fields, describing besides gravitons and
photons also additional vector boson fields and scalars.\\

Let us study some of the even ``basis vectors'', representing 
${}^{I}\hat{\cal A}^{m \dagger}_{f}$ and 
${}^{II}\hat{\cal A}^{m \dagger}_{f}$, looking for them either as 
algebraic products of fermions and their Hermitian conjugated partners, 
or by using Eqs.~(\ref{calIAb1234gen}, \ref{calbIIA1234gen}).

One can find ${}^{I}\hat{\cal A}^{2 \dagger}_{3}$ by the algebraic 
product of $\hat{b}^{2 \dagger }_{1} \,*_{A}\,
(\hat{b}^{1 \dagger }_{1} )^{\dagger} $:
$$\hat{b}^{2 \dagger }_{1} (\equiv \stackrel{03}{[-i]}\stackrel{12}{(-)}
\stackrel{56}{[+]}) \,*_{A}\,(\hat{b}^{1 \dagger }_{1} )^{\dagger}
(\equiv \stackrel{03}{(+i)}\stackrel{12}{[+]}\stackrel{56}{[+]})^{\dagger}
\rightarrow {}^{I}\hat{\cal A}^{2 \dagger}_{3}(\equiv \stackrel{03}{(-i)}
\stackrel{12}{(-)}\stackrel{56}{[+]},$$ or by looking for
${}^{I}\hat{\cal A}^{m \dagger}_{f}$, which applying on
$\hat{b}^{1 \dagger }_{1} $ transforms it to $\hat{b}^{2 \dagger }_{1}$:
$${}^{I}\hat{\cal A}^{m \dagger}_{f}(\equiv \stackrel{03}{(-i)}
\stackrel{12}{(-)}\stackrel{56}{[+]})\,*_{A}\,\hat{b}^{1 \dagger }_{1}
(\equiv \stackrel{03}{(+i)}\stackrel{12}{[+]}\stackrel{56}{[+]})
\rightarrow \hat{b}^{2 \dagger }_{1} (\equiv \stackrel{03}{[-i]}
\stackrel{12}{(-)}\stackrel{56}{[+]}).$$

This ${}^{I}\hat{\cal A}^{m \dagger}_{f}(\equiv \stackrel{03}{(-i)}
\stackrel{12}{(-)}\stackrel{56}{[+]})={}^{I}\hat{\cal A}^{2 \dagger}_{3}$,
transforms the fermion of right-handedness with spin up to the fermion of 
right-handedness with spin down. {\bf We recognise it as the even ``basis vector''
of graviton} (which in tensor product with the basis in ordinary space-time
and carrying the space index $\mu$ presents the graviton). In
Table~\ref{Table Clifffourplet.} is placed on the second line of the third
column.

The even ``basis vector'' of the graviton which transforms 
$\hat{b}^{2 \dagger }_{1}$ into  
$\hat{b}^{1 \dagger }_{1}$ is ${}^{I}\hat{\cal A}^{1 \dagger}_{4}
(\equiv \stackrel{03}{(+i)}\stackrel{12}{(+)}\stackrel{56}{[+]})$, 
appearing in the first line of the fourth column.

The even ``basis vector'' of the {\bf graviton in $d=(13+1)$}, which would 
transform the right-handed electron with spin up into the right-handed
electron with spin down, presented in Table~6 of the Ref.~\cite{n2025Bled}~
on the 27 and 28 lines, would have a similar construction as 
${}^{I}\hat{\cal A}^{2 \dagger}_{3}$, namely 
${}^{I}\hat{\cal A}^{\dagger}_{e^-_{R\uparrow} \rightarrow 
e^{-}_{R\downarrow}} (\equiv \stackrel{03}{(-i)}\stackrel{12}{(-)}
\stackrel{56}{[-]}\stackrel{78}{[-]}\stackrel{9\,10}{[+]}$
$\stackrel{11\,12}{[+]}\stackrel{13\,14}{[+]}$; all the eigenvalues of the 
Cartan subalgebra members except $\stackrel{03}{(-i)}
\stackrel{12}{(-)}$ must be zero, that means that the only nilpotents 
must appear in the first two columns, all the rest must be projectors.\\

The even ``basis vectors'' representing the internal space of photons, 
having no charges, must be constructed from only projectors, either in the 
internal space of $d=(5+1)$, or in the internal space of $d=(13+1)$. \\

\vspace{2mm}

Let us generate some of the even ``basis vectors'' of the second group 
${}^{II}\hat{\cal A}^{m \dagger}_{f}$, presented at 
Table~\ref{Table Clifffourplet.} in the last four columns. We can do this 
with the algebraic products of Hermitian conjugated partners of the 
even ``basis vectors'' and the even ``basis vectors'', 
Eq.~(\ref{AIIbdaggerb}), or by using Eq.~(\ref{calbIIA1234gen}).

 $${}^{II}\hat{\cal A}^{1 \dagger}_{3}  (\equiv \stackrel{03}{[-i]}
 \stackrel{12}{[+]}\stackrel{56}{[+]}) = 
 (\hat{b}^{1 \dagger }_{1})^{\dagger}\,*_A\,\hat{b}^{1 \dagger }_{1}$$.

 Eq.~(\ref{calbIIA1234gen}) requires: 
$$\hat{b}^{1 \dagger }_{1}
(\equiv \stackrel{03}{(+i)}\stackrel{12}{[+]}\stackrel{56}{[+]})\,*_A\,
{}^{II}\hat{\cal A}^{1 \dagger}_{3} (\equiv \stackrel{03}{[-i]}
 \stackrel{12}{[+]}\stackrel{56}{[+]}) \rightarrow \hat{b}^{1 \dagger }_{1}$$.
 
 Let be added that ${}^{II}\hat{\cal A}^{1 \dagger}_{3}=$
 $(\hat{b}^{2 \dagger }_{1})^{\dagger}\,*_A\,\hat{b}^{2 \dagger }_{1}=$
  $(\hat{b}^{m \dagger }_{1})^{\dagger}\,*_A\,\hat{b}^{m \dagger }_{1}$, 
for all $m=(1,2,3,4)$ of the first family.\\

One can check that the same is true also for all the members of Table~6 of 
Ref.~\cite{n2025Bled}; Any of the $64$ members, either quarks or leptons, 
as well as antiquarks and antileptons of this family generates 
the same ${}^{II}\hat{\cal A}^{\dagger}_{e^-_{R\uparrow} \rightarrow 
e^{-}_{R\uparrow}} $
$${}^{II}\hat{\cal A}^{\dagger}_{e^-_{R\uparrow} \rightarrow 
e^{-}_{R\uparrow}} (\equiv \stackrel{03}{[-]}\stackrel{12}{[+]}|
\stackrel{56}{[+]}\stackrel{78}{[-]}||\stackrel{9\,10}{[-]}
\stackrel{11\,12}{[-]}\stackrel{13\,14}{[-]} =
(e^-_{R\uparrow})^{\dagger}  (\equiv \stackrel{03}{(+i)}\stackrel{12}{[+]}|
\stackrel{56}{(-)}\stackrel{78}{[-]}|| \stackrel{9\,10}{(+)}
\stackrel{11\,12}{(+)}\stackrel{13\,14}{(+)})^{\dagger}\,*_A\, 
e^-_{R\uparrow}$$.




%
\subsection{Feynman diagrams in our way and questions to be answered}
%
\label{Feynmanoursandrest}

The action for fermion and boson second quantised fields, Eq.~(\ref{action}),
demonstrating the relations among fermion and boson ``basis vectors'',
presented in Eqs.~(\ref{AIAIIorth} - \ref{AIIbdaggerb}), determines
Feynman diagrams for our description of internal spaces.\\

Let us shortly repeat the differences between our way of describing the
internal spaces of fermion and boson fields, and the usual way - the
most noticeable differences:\\
{\it a.\,\,\,} The odd (anti-commuting) ``basis vectors'', describing the
internal spaces of fermion fields appear in families ($2^{\frac{d}{2}-1}$);
In ordinary theories, the families are postulated, and the anti-commutativity is postulated; the internal spaces of fermions are described by matrices
in fundamental representations;\\
{\it b.\,\,} Each family  (with $2^{\frac{d}{2}-1}$  members) contains
in $d=2(2n+1)$ ``basis vectors'' of fermions and anti-fermions, the
Hermitian conjugate partner of the odd ``basis vectors'' of fermions
appear in a separate group, no Dirac sea is correspondingly needed,
as we see in Table~\ref{Table Clifffourplet.} for $d=(5+1)$, and in
Table~2 in the reference~\cite{n2025Bled} for $d=(13+1)$;
In ordinary theories, the antifermions are postulated as the holes in
the Dirac sea; The interpretation of a particle-antiparticle pair as
the particle taken out of the Dirac sea, while a missing particle in
the Dirac sea is interpreted as an antiparticle, requires that a
particle-antiparticle annihilation is interpreted as the particle going
back to the Dirac sea;\\
{\it c.\,\,} The algebraic  products of odd ``basis vectors'', independently 
to which family they belong, are mutually orthogonal, 
Eq.~(\ref{orthogonalodd}), and so are mutually orthogonal also their 
Hermitian conjugate partners; The algebraic products of the odd 
``basis vectors'' with their Hermitian conjugate partners are non-zero;
\begin{eqnarray}
\hat{b}^{m \dagger}_f \,*_{A}\, \hat{b}^{m `\dagger }_{f }&=& 0\,,
\quad \hat{b}^{m}_f \,*_{A}\, \hat{b}^{m ` }_{f }= 0\,,\nonumber\\
&&
\forall \, (m,m',f)\,,\nonumber\\
\hat{b}^{m }_f *_{A} \hat{b}^{m `\dagger }_{f }&\ne& 0\,;
\label{nonorthogonalodd}
\end{eqnarray}
%


However, in the case $d=4n$, the families include only fermions, no 
antifermions. In this case the Dirac sea might help. Namely, if 
we choose the appropriate families, the Hermitian conjugate values 
of charges of the odd ``basis vectors'' can have the opposite values 
for charges as the ``basis vectors''. Let us treat the case SO(7,1), 
choosing the families, so that the Hermitian conjugate partners carry the 
opposite charge. It is not difficult to continue this
 Eq.~(\ref{allcartaneigenvec4nnice})  with the choices of appropriate 
families for the remaining four cases. However, this construction, 
jumping among different families, is unacceptable. A better 
advice is to enlarge the internal space to $d=2(2n+1)$.
\begin{small}
\begin{eqnarray}
\label{allcartaneigenvec4nnice}
&& \qquad \qquad \qquad \qquad d=4n\, ,\nonumber\\
&& \hat{b}^{1 \dagger}_{2}=\stackrel{03}{(+i)}\stackrel{12}{[+]}
\stackrel{5 6}{(+)} \stackrel{7 8}{(+)}\,,\qquad  
\hat{b}^{1}_{2}=\stackrel{03}{(-i)}\stackrel{12}{[+]}
\stackrel{5 6}{(-)} \stackrel{7 8}{(-)}\,
\nonumber\\
&&\hat{b}^{2 \dagger}_{2} = \stackrel{03}{[-i]} \stackrel{12}{(-)}
\stackrel{56}{(+)}\stackrel{78}{(+)}\,,\qquad \,
\hat{b}^{2 }_{2} = \stackrel{03}{[-i]} \stackrel{12}{(+)}
\stackrel{56}{(-)} \stackrel{78}{(-)}\,,
\nonumber\\ 
&&\hat{b}^{3 \dagger}_{1} = \stackrel{03}{(+i)}\stackrel{12}{[+]}
\stackrel{56}{(-)} \stackrel{78}{(-)}\,, \quad\,\,\,\,\,\,
\hat{b}^{3}_{1} =  \stackrel{03}{(-i)}\stackrel{12}{[+]}
\stackrel{56}{(+)} \stackrel{78}{(+)}\,\,,
 \nonumber\\
 &&\hat{b}^{4 \dagger}_{1} = \stackrel{03}{[-i]} \stackrel{12}{(-)}
 \stackrel{56}{(-)} \stackrel{78}{(-)}\,, \quad\,\,\,\,\,\,
\hat{b}^{4}_{1} =   \stackrel{03}{[-i]} \stackrel{12}{(+)}
\stackrel{56}{(+)} \stackrel{78}{(+)}\,\,,
%
\end{eqnarray}
\end{small}
(In addition, this construction limits the number of families to only one family. 
Correspondingly, the families must be postulated ``by hand''.)\\ 
{\it d.\,\,} The commuting ``basis vectors'', describing  the internal spaces 
of boson fields appear in two orthogonal groups, having their Hermitian 
conjugated partners within each group; The ordinary theories recognise 
only one kind of fields (although the scalar fields might be recognised 
as the second kind), the commutativity is postulated;\\
{\it e.\,\,} Both commuting ``basis vectors'' are expressible by algebraic 
products of odd ``basis vectors'' and their Hermitian conjugated 
partners, Eqs.~(\ref{AIbbdagger}, \ref{AIIbdaggerb}); Ordinary 
theories describe internal spaces  of bosons with matrices in the adjoint 
representations;\\

 {\it The differences in the description of the internal spaces of 
fermion and boson fields in our case, and in usual cases, cause the
differences in presenting  Feynman diagrams.}\\ 

 The most noticeable difference is that our description of the 
internal spaces of fermion fields tells us that all the odd ``basis vectors'' 
are mutually orthogonal, Eq.~(\ref{orthogonalodd}), and so are mutually 
orthogonal also their Hermitian conjugate partners. We expect that our 
Feynman diagrams will differ from the usual ones when fermions and 
antifermion meet.

In our case, the two odd ``basis vectors'' can interact only by exchanging 
a boson represented with the even ``basis vectors'', as demonstrated 
in Eqs.(\ref{calIAb1234gen}, \ref{calbIIA1234gen}). The particle in 
ordinary theories (leaving the hole in the Dirac sea) resembles our 
particle (except that our particles are massless, and have their internal 
space presented by odd ``basis vectors'', and not by matrices), while 
the antiparticle (in ordinary theories, its hole in the Dirac sea), does not 
really resemble our antiparticle (of opposite charges to the particles,
belonging to the same family~\cite{n2025Bled}, unless the break of 
symmetry mixes families, bringing them masses. Our antiparticles move
in the same way as particles; this is not the case with the hole in the 
Dirac sea.

Let us start with drawing the Feynman diagram for a fermion, representing
an electron, with the internal space described by the odd ``basis vector''
$\hat{b}^{1\dagger }_1 (\equiv \stackrel{03}{(+i)}\stackrel{12}{[+]}
\stackrel{56}{[+]})$, Table~\ref{Table Clifffourplet.}, with the momentum
$\vec{p}_1$, radiating the photon with the even ``basis vector''
${}^{II}\hat{\cal A}^{1 \dagger}_{3} (\equiv \stackrel{03}{[-i]} 
\stackrel{12}{[+]}\stackrel{56}{[+]}\equiv
 (\hat{b}^{1 \dagger }_{1})^{\dagger}\, *_{A}\,
\hat{b}^{1 \dagger }_{1}$) with the momentum  $\vec{p}_3 $, 
while the electron with the odd ``basis vector'' $\hat{b}^{1 \dagger }_{1}$  
; $\hat{b}^{1\dagger }_1 (\equiv \stackrel{03}{(+i)}\stackrel{12}{[+]}
\stackrel{56}{[+]})\,*_A\,{}^{II}\hat{\cal A}^{1 \dagger}_{3} 
(\equiv \stackrel{03}{[-i]} \stackrel{12}{[+]}\stackrel{56}{[+]}) \rightarrow 
\hat{b}^{1 \dagger }_{1}$; continues its way with smaller momentum  
$\vec{p}_2$.
%
This event, when an electron radiates a photon, is presented in Fig.~\ref{bIIAb}.
The equivalent diagram is valid also for the electron in the ordinary theories,
only the photon will not be described in our way.
\\
\begin{small}
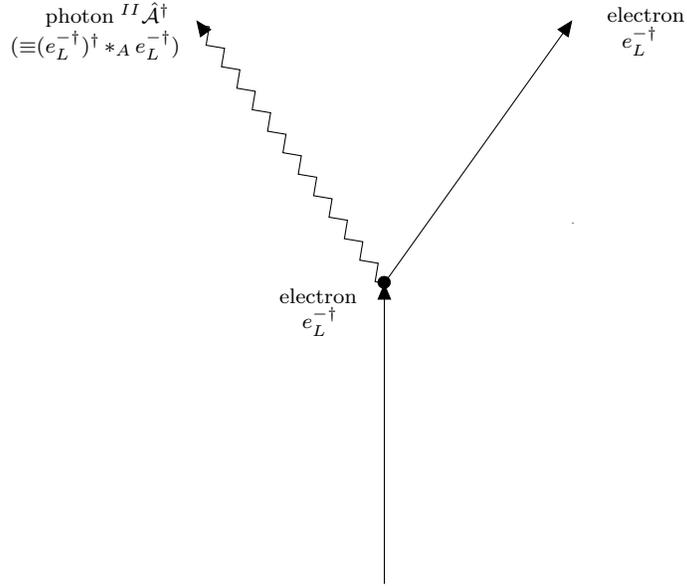
\begin{figure}
\centering
\begin{tikzpicture}[>=triangle 45]
\draw [->]
(1,1.5) node {$\bullet$}-- (3.5,5) 
node [anchor=west]
{$
\begin{smallmatrix} \\
\\
\;\;\;{\rm electron} \;
\\
e^{-\dagger}_{L}
\end{smallmatrix} $};
\draw [-]
(1,1.5) node {$\bullet$} -- (1,1.5) node {$\bullet$};) 
\draw [->,snake]
(1,1.5) node {$\bullet$}-- (-1.5,5)
node [anchor=east]
{$
\begin{smallmatrix} \\
\\
\;\;\;\;{\rm photon}\;
{}^{II}\hat{{\cal A}}^{\dagger}
\\(\equiv (e^{-\dagger}_{L})^{\dagger}\,*_{A}\,e^{-\dagger}_{L})
\end{smallmatrix} 
$}; 
\draw [->]
(1,-2.5) node {$$}-- (1,1.5) 
node [anchor=east]
{$
\begin{smallmatrix} \\ 
\\
\\ 
\\
{\rm electron}\,\,\,\\
e^{- \dagger}_{L} \,\,
\end{smallmatrix} $}; 
\draw [ ] 
(3.5,2.3) node {$$}-- (3.5,2.3)
node [anchor=east]
{$
$}; 
%
\end{tikzpicture}
  \caption{\label{bIIAb} 
An electron, with the internal space described by 
$\hat{b}^{1 \dagger}_{1}$ and with the momentum $\vec{p}_1$ in 
ordinary space, Table~\ref{Table Clifffourplet.}, radiates 
the photon with the ``basis vector'' ${}^{II}\hat{\cal A}^{1 \dagger}_{3}
(\equiv \stackrel{03}{[-i]} \stackrel{12}{[+]}\stackrel{56}{[+]}\equiv
(\hat{b}^{1 \dagger }_{1})^{\dagger}\, *_{A}\,
\hat{b}^{1 \dagger }_{1}$), with the momentum  $\vec{p}_3 $,  
   while the electron, $\hat{b}^{1 \dagger }_{1}$,  continues its way 
 with a smaller momentum  $\vec{p}_2$, Fig.~\ref{bIIAb}. This diagram 
 is representing also the electron in the usual theories, except that photons  
 are not presented in our way. For the electron 
 with the ``basis vector'', $e^{-\dagger}_{L} (\equiv \stackrel{03}{[- i]} 
 \stackrel{12}{[+]} \stackrel{56}{(-)} \stackrel{78}{(+)} 
 \stackrel{9\, 10}{(+)} \stackrel{11\,12}{(+)} \stackrel{13\,14}{(+)})$, 
 from Table~2 in Ref.~\cite{n2025Bled}, the photon with the ``basis 
 vector''  ${}^{II}\hat{\cal A}^{\dagger}(\equiv (e^{-\dagger}_{L})^{\dagger}$
 $ \, *_A\, e^{-\dagger}_{L} \equiv \stackrel{03}{[- i]} \stackrel{12}{[+]} $
 $ \stackrel{56}{[+]} \stackrel{78}{[-]} \stackrel{9\, 10}{[-]}
 \stackrel{11\,12}{[-]} \stackrel{13\,14}{[-]})$ takes away the momentum.
}.
\end{figure}
\end{small}
%
The equivalent Feynman diagram represents, in the case that the odd ``basis vectors" describe the internal space of fermions, also the event
that a positron, with the internal space described by
$\hat{b}^{3 \dagger}_1 (\equiv \stackrel{03}{[-i]}\stackrel{12}{[+]}
\stackrel{56}{(-)})$, Table~\ref{Table Clifffourplet.}, and with the
momentum $\vec{p}_1$ in ordinary space, radiates a photon with the
internal space represented by the even ``basis vector''
${}^{II}\hat{\cal A}^{1 \dagger}_{3} (\equiv \stackrel{03}{[-i]}
\stackrel{12}{[+]}\stackrel{56}{[+]}\equiv
(\hat{b}^{3 \dagger }_{1})^{\dagger}\, *_{A}\,
\hat{b}^{3 \dagger }_{1}$). Either the electron or the positron belongs
to the same family. (Their algebraic products are zero.) However, the corresponding Feynman diagram in the usual theories,
representing the positron, should have the arrows for the positron turned back, 
$\uparrow$ should be turned into $\downarrow$.
 
The event, when a positron $\hat{b}^{3 \dagger }_{1}$
radiates a photon  ${}^{II}\hat{\cal A}^{1 \dagger}_{3}$, is presented in Fig.~\ref{bIIAbpos}.\\
%
\begin{small} 
\begin{figure}
\centering
\begin{tikzpicture}[>=triangle 45]
\draw [->]
(1,1.5) node {$\bullet$}-- (3.5,5) 
node [anchor=west]
{$
\begin{smallmatrix} \\
\\
\;\;\;{\rm positron} \;
\\
e^{+\dagger}_{R}
\end{smallmatrix} $};
\draw [-]
(1,1.5) node {$\bullet$} -- (1,1.5) node {$\bullet$};) 
\draw [->,snake]
(1,1.5) node {$\bullet$}-- (-1.5,5)
node [anchor=east]
{$
\begin{smallmatrix} \\
\\
\;\;\;\;{\rm photon}\;
{}^{II}\hat{{\cal A}}^{\dagger}
\\(\equiv (e^{+\dagger}_{R})^{\dagger}\,*_{A}\,e^{+\dagger}_{R})
\end{smallmatrix} 
$}; 
\draw [->]
(1,-2.5) node {$$}-- (1,1.5) 
node [anchor=east]
{$
\begin{smallmatrix} \\ 
\\
\\ 
\\
{\rm positron}\,\,\,\\
e^{+ \dagger}_{R} \,\,
\end{smallmatrix} $}; 
\draw [ ] 
(3.5,2.3) node {$$}-- (3.5,2.3)
node [anchor=east]
{$
$}; 
%
\end{tikzpicture}
%
  \caption{\label{bIIAbpos}   
A positron, with the internal space described by 
$\hat{b}^{3 \dagger}_{1}$ and with the momentum $\vec{p}_1$ in 
ordinary space, Table~\ref{Table Clifffourplet.}, radiates 
the photon with the ``basis vector'' ${}^{II}\hat{\cal A}^{1 \dagger}_{3}
(\equiv \stackrel{03}{[-i]} \stackrel{12}{[+]}\stackrel{56}{[+]}\equiv
(\hat{b}^{3 \dagger }_{1})^{\dagger}\, *_{A}\,
\hat{b}^{3 \dagger }_{1}=(\hat{b}^{1 \dagger }_{1})^{\dagger}\, *_{A}\,
\hat{b}^{1 \dagger }_{1}$ 
), with the momentum  $\vec{p}_3 $,  
   while the positron, $\hat{b}^{3 \dagger }_{1}$,  continues its way 
 with smaller momentum  $\vec{p}_2$, Fig.~\ref{bIIAbpos}. 
For the positron  with the ``basis vector'', $e^{+\dagger}_{R} 
(\equiv \stackrel{03}{(+i)}  \stackrel{12}{[+]} \stackrel{56}{[+]} 
\stackrel{78}{[-]} \stackrel{9\, 10}{[-]} \stackrel{11\,12}{[-]}
 \stackrel{13\,14}{[-]})$,  from Table~2 in Ref.~\cite{n2025Bled}, the 
 photon with the ``basis  vector''  ${}^{II}\hat{\cal A}^{\dagger} 
 (\equiv (e^{+\dagger}_{R})^{\dagger}$
 $ \, *_A\, e^{+\dagger}_{R} \equiv \stackrel{03}{[- i]} \stackrel{12}{[+]} $
 $ \stackrel{56}{[+]} \stackrel{78}{[-]} \stackrel{9\, 10}{[-]}
 \stackrel{11\,12}{[-]} \stackrel{13\,14}{[-]})$ takes away the momentum.
 The corresponding Feynman diagram in the usual theories,
representing the positron should have the arrows for the positron turned back, 
$\uparrow$ should be turned into $\downarrow$. 
}.
\end{figure}
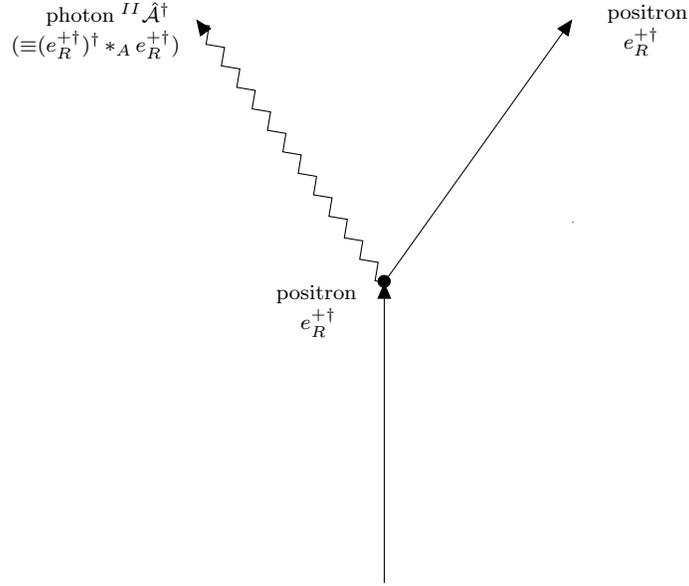
\end{small}

Since the fermions in our case differ from fermions in the usual theory -
our fermions and antifermions belong to the same family, while the
families distinguish among themselves only in the family quantum numbers - let us see how we can draw the Feynman rule for the annihilation of an
electron and positron in the case that the internal space has $d=2(2n+1)$,
the most promising is $d=(13+1)$ dimensions, this choice offers all the
quarks and leptons and antiquarks and antileptons, observed at low
energies in an elegant way, treating all the boson fields in an equivalent
way, with gravitons included. 
Looking at the figure~\ref{Figepnew}, we see that it differs from the
corresponding Feynman diagram for the electron-positron annihilation in
usual theories:  In our case, the electron, after radiating a photon
${}^{II}\hat{{\cal A}}^{\dagger}$, continues its way to the right, up to the
positron, which is coming up and after radiating the photon, turns to the left. 
They remain in a vacuum together without the momentum in ordinary space.
In standard theories, the positron goes down rather than up.

Taking into account figures~\ref{bIIAb} and~\ref{bIIAbpos}, to try to make the diagram
 as close to the usual diagrams as possible, lead to Fig.~\ref{Figepnew}.
\begin{small} 
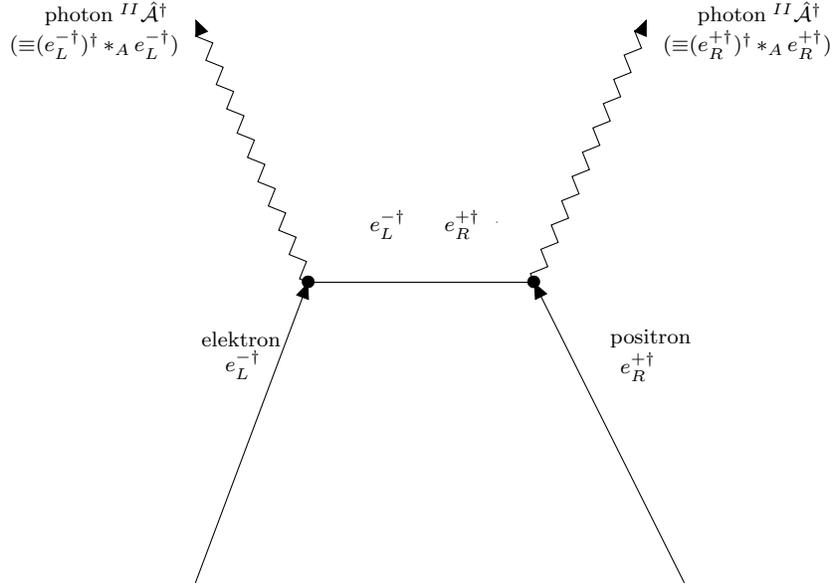
\begin{figure}
  \centering
  \begin{tikzpicture}[>=triangle 45]
\draw [->,snake]
 (4,1.5) node {$\bullet$}-- (5.5,5)
node [anchor=west]
{$
\begin{smallmatrix} \\
\\
\;\;\;\;{\rm photon}\;
{}^{II}\hat{{\cal A}}^{\dagger}
\\(\equiv (e^{+\dagger}_{R})^{\dagger}\,*_{A}\,e^{+\dagger}_{R})
\end{smallmatrix} 
$}; 
\draw [-]
(1,1.5)  node {$\bullet$} -- (4,1.5) node {$\bullet$};)
%
\draw [->,snake]
 (1,1.5) node {$\bullet$}-- (-0.5,5)
node [anchor=east]
{$
\begin{smallmatrix} \\
\\
\;\;\;\;{\rm photon}\;
{}^{II}\hat{{\cal A}}^{\dagger}
\\(\equiv (e^{-\dagger}_{L})^{\dagger}\,*_{A}\,e^{-\dagger}_{L})
\end{smallmatrix} 
$};  
  \draw [->]                                   
(6,-2.5) node {$$}-- (4,1.5)
node [anchor=west]
{$\;\;\;\;
\begin{smallmatrix} \\
\\
\\
\\
\\
\\
\\
\\
\\                               
                                      \;\;\;\;{\rm positron}\\
                                      e^{+ \dagger}_{R}
                                      \end{smallmatrix} $}; 
   \draw [->]                                   
(-0.5,-2.5) node {$$}-- (1,1.5)
node [anchor=east]
{$
\begin{smallmatrix} \\
\\
\\
\\
\\
\\
\\
\\ 
\\
                                          {\rm elektron}\,\,\,\\
                                          e^{- \dagger}_{L}  \,\,                                     
                                      \end{smallmatrix} $}; 
  \draw [ ] 
(3.5,2.3) node {$$}-- (3.5,2.3) 
  node [anchor=east] 
{$
\begin{smallmatrix} \\
                                 \; e^{- \dagger}_{L}\,\, 
                                 \;\; \;\;\; e^{+ \dagger}_{R}
                                      \end{smallmatrix} $}; 
%
\end{tikzpicture}
  \caption{\label{Figepnew} 
The left-hand side represents the path of the electron, $e^{-\dagger}_{L}$,
which radiates a photon   $(e^{-\dagger}_{L})^{\dagger}\,*_{A}\,
e^{-\dagger}_{L}$, and continues its way straight to the right, up to a positron,
$e^{+\dagger}_{R}$ coming up. They both  radiate a photon
$(e^{-\dagger}_{L})^{\dagger}\,*_{A}\,
e^{-\dagger}_{L}$ and $(e^{+\dagger}_{R})^{\dagger}\,*_{A}\,
e^{+\dagger}_{R}$ (both are of the same kind) and remain
without momenta in the quantum vacuum. It can also happen the opposite:
The positron, $e^{+\dagger}_{R}$, radiates a photon
$(e^{+\dagger}_{R})^{\dagger}\,*_{A}\,e^{+\dagger}_{R}$, and
continues its way straight to the left, up to an electron,
$e^{-\dagger}_{L}$ coming up from the left hand side. Both  radiate a photon
$(e^{-\dagger}_{L})^{\dagger}\,*_{A}\,e^{-\dagger}_{L}$ of the same kind.
Both remain without momentum in the quantum vacuum.
}.
\end{figure}
\end{small}
Although the Feynman diagram for the electron-positron annihilation, 
presented in Fig.~\ref{Figepnew}, seems quite close to what we 
are looking for, it leaves open the question whether the electron 
and positron transfer all the momentum to the two photons.\\

Let us try with a slightly different interpretation. 
The electron $e^{-\dagger}_{L}$
radiates a photon $(e^{-\dagger}_{L})^{\dagger}\,*_{A}\, e^{-\dagger}_{L}$,
turns to the right and meets a positron $e^{+\dagger}_{R}$ who already emitted
a photon  $(e^{+\dagger}_{R})^{\dagger}\,*_{A}\,e^{+\dagger}_{R}$, and
has turned to the left. They go together into the quantum vacuum without the
momenta in ordinary space-time, as presented in Fig.~\ref{Hepann}.
%

\begin{small} 
\begin{figure}
  \centering
  \begin{tikzpicture}[>=triangle 45]
\draw [->,snake]
(4,1.5) node {$\bullet$}-- (5.5,5)
node [anchor=west]
{$
\begin{smallmatrix} \\
\\
\;\;\;\;{\rm photon}\;
{}^{II}\hat{{\cal A}}^{\dagger}
\\(\equiv (e^{+\dagger}_{R})^{\dagger}\,*_{A}\,e^{+\dagger}_{R})
\end{smallmatrix} 
$}; 
\draw (2.5,1.5) circle (1.0);
\draw [->]
 (1,1.5)  node {$\bullet$} -- (1.5,1.5) 
 ;)
\draw [<-]
(3.5,1.5) 
 -- (4,1.5) node {$\bullet$};)
\draw [->,snake]
 (1,1.5) node {$\bullet$}-- (-0.5,5)
node [anchor=east]
{$
\begin{smallmatrix} \\
\\
\;\;\;\;{\rm photon}\;
{}^{II}\hat{{\cal A}}^{\dagger}
\\(\equiv (e^{-\dagger}_{L})^{\dagger}\,*_{A}\,e^{-\dagger}_{L})
\end{smallmatrix} 
$};  
  \draw [->]                                   
(6,-2.5) node {$$}-- (4,1.5)
node [anchor=west]
{$\;\;\;\;
\begin{smallmatrix} \\
\\
\\
\\
\\
\\
\\
\\
\\                               
                                      \;\;\;\;{\rm positron}\\
                                      e^{+ \dagger}_{R}
                                      \end{smallmatrix} $}; 
   \draw [->]                                   
(-0.5,-2.5) node {$$}-- (1,1.5)
node [anchor=east]
{$
\begin{smallmatrix} \\
\\
\\
\\
\\
\\
\\
\\ 
\\
                                          {\rm elektron}\,\,\,\,\,\,\\
                                           {\rm elektron}\,\,\,\,\,\,\\
                                     \end{smallmatrix} $}; 
  \draw [ ] 
(3.5,2.3) node {$$}-- (3.5,2.3) 
  node [anchor=east] 
{$
\begin{smallmatrix} \\  \;\; \;\;\;                                
   e^{- \dagger}_{L}\,\, 
                            \;\; \;\;\;   \;\; \;\;\;      \;\; \;\;\; e^{+ \dagger}_{R}\\
                            \\
                            \\
                            \\
                            \\ 
                                     \;\;\;\;{\rm vacuum}\;
                                      \end{smallmatrix} $}; 
{$
$}; 

%
\end{tikzpicture}
  \caption{\label{Hepann} The electron $e^{- \dagger}_L$ radiates
  a photon  ${}^{I}\hat{{\cal A}}^{\dagger}_{phee^{\dagger}}
  (\equiv e^{-\dagger}_{L}\,*_{A}\,(e^{-\dagger}_{L})^{\dagger})$, and 
  goes to the right to the vacuum.  The positron $e^{+ \dagger}_L$, radiates
  a photon  ${}^{I}\hat{{\cal A}}^{\dagger}_{phpp^{\dagger}}
  (\equiv e^{+\dagger}_{R}\,*_{A}\,(e^{+\dagger}_{R})^{\dagger}$,
   and turning to the left remains with electron in the vacuum.} 
\end{figure}
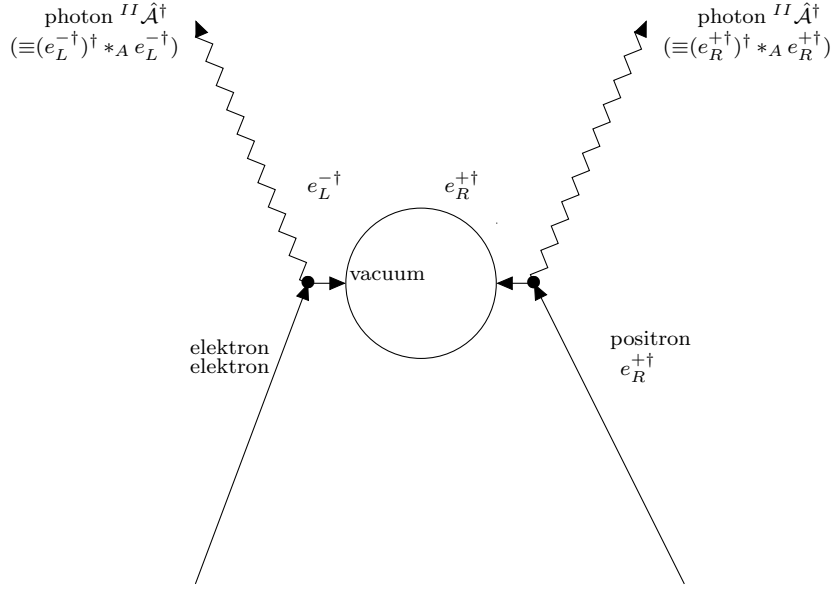
\end{small}
%

The symbolic diagram with the electron and the positron going into the
vacuum ``simultaneously'' is of course expected to be/become the usual
propagator for a fermion.

We might argue for that by writing down the properties which this
propagator-like operator must have with respect to symmetries (the
vacuum must remain symmetric under the symmetries of the
theory) and the properties of causality as to which particle is to
propagate only forward in time.

In the present article, we shall postpone these arguments for getting the
usual propagator, but it is, of course, logically needed to argue for it.
If one assumes the usual Dirac sea vacuum, it should be rather obvious
what our line with the vacuum blob in the middle must be.

We need in the next step to present all the measured Feynman diagrams in our way; that is, with fermions (quarks and leptons and antiquarks and antileptons), whose internal space is
described by the odd ``basis vectors'', the  ``basis vectors'' with the odd number of nilpotents, which are all mutually orthogonal, and with bosons (gravitons, weak bosons, photons, gluons, scalars), whose internal spaces are described by the even "basis vectors '' the ``basis vectors'' with the even number of nilpotents, which appear in two
orthogonal groups. We must see whether we can agree with the experiments and find
a way to represent them that we will agree on.

\section{Presenting open problems concerning Feynman diagrams}
\label{openproblems}

Accepting the idea of the papers~\cite{n2023NPB,n2024NPB,n2023MDPI,n2024IARD}
that {\it the internal spaces (spins and charges) of fermions and bosons are
described by ``basis vectors'' which are the odd } (for fermions~%
\cite{nh2021RPPNP}) {\it and even} (for bosons) {\it products of nilpotents},
Eq.~(\ref{nilproj}), the authors are trying to find out whether and
up to what extent ``nature manifests'' the proposed idea,
offering hopefully the unifying theory of gravity, all the gauge fields, the scalar fields, and
the fermion and antifermion fields.

In this contribution, we study {\it massless fermion and boson fields under the
condition that they have non-zero momentum only in $d=(3+1)$, while
internal spaces have $d=2(2n+1)$}, the choice of $d=(13+1)$ offers the
description of the second quantised quarks, leptons and antiquarks and
antileptons and of all the second quantised vector (gravitons, weak bosons,
photons, gluons) and scalar fields.

Let us mention again that if we choose the internal space with $d=4n$, that
is $d=(4n-1) +1$, the families include only fermions, no antifermions; Eq.~(\ref{allcartaneigenvec4n}) manifests the properties of the corresponding
``basis vectors'' in the case that $4n=7+1$. (In such cases, the Dirac sea
would be needed. The more elegant choice is to enlarge the internal space
to $d=4n+2$, as it is $d=(13+1)$, which offer the description that quarks
and leptons distinguish only in the $SO(6)$ part of $SO(13+1)$, and antiquarks
and antileptons distinguish only in the $SO(6)$ part of $SO(13+1)$.)

All the fields are tensor products of the odd (fermion fields) and even 
(boson fields) ``basis vectors'' and basis in ordinary space-time, while the boson
fields have in addition the space index $\alpha$ ($\alpha =\mu=(0,1,2,3)$
for vectors, and $\alpha =\sigma \ge 5)$ for scalars).  We correspondingly
have the Poincaré symmetry only in $d=(3+1)$.

The algebraic products of ``basis vectors'' of boson and fermion fields
determine the action for fermions and bosons, Eqs.~(\ref{orthogonalodd}-
\ref{wholespacebosons}).

The odd (anti-commuting) ``basis vectors'',  appear in families,  including in
$d=2(2n+1)$ fermions and antifermions (all odd ``basis vectors'' are
mutually orthogonal, Hermitian conjugate partner of the odd ``basis
vectors'' appear in a separate group, no Dirac sea is correspondingly
needed);
In ordinary theories, the families are postulated, and the anti-commutativity 
is postulated; matrices describe the internal spaces of fermions in 
fundamental representations; The antifermions are postulated as the holes in 
the Dirac sea.

The even (commuting) ``basis vectors'',  appear in the proposed theory
in two orthogonal groups; and all even ``basis vectors'' are expressible
by algebraic products of odd ``basis vectors'' and their Hermitian
conjugate partners.
In ordinary theories, instead of our even ``basis vectors'' the matrices
in adjoint representations are used.

The difference in properties of the second-quantised fields in the proposed
theory and the ordinary theories require, among many other things, studying
also the Feynman diagrams and compare them to the experimentally
confirmed the Feynman diagrams of the ordinary theories.

The symbolic diagram with the electron and the positron going into the
vacuum ``simultaneously'' is expected to become the usual
propagator for a fermion.

In the present article, we postponed the arguments about the properties
which the propagator-like operator must have with respect to symmetries
of the theory and the properties of causality.


We need to present all the measured Feynman diagrams in our way; that is,
with fermions (quarks and leptons, antiquarks and antileptons), whose
internal space is described by the odd ``basis vectors'', which are all
mutually orthogonal, and
with bosons (gravitons, weak bosons, photons, gluons, scalars), whose
internal space is described by the even "basis vectors'' which appear in two
orthogonal groups. We expect that we can agree with the
experiments and find
a way to represent them that we will agree on.

Let us conclude by saying that if we describe the internal spaces of fermions
and bosons with the  "basis vectors'' in $d=(13+1)$, and assume that
fermion and boson fields have non-zero momentum only in $d=(3+1)$ of
the ordinary space-time, then we unify gravity and all the gauge fields:
$SO(3,1)$ determines spins and handedness of gravitons, fermions, and
antifermions, $SU(2)\times SU(2)$ determine weak charges of fermions
and bosons, $SU(3)\times U(1)$ determine the colour charges of quarks
and antiquarks and gluons. Photons' "basis vectors'' are products of
only projectors, with all spins and charges equal to zero, gravitons'
"basis vectors'' have two nilpotents only in $SO(3,1)$ part of
$SO(13,1)$, weak bosons' "basis vectors'' have two nilpotents in
$SU(2)$ part of $SO(13,1)$, gluons' "basis vectors'' have two nilpotents
in $SO(6)$ part of $SO(13,1)$. Fermions' "basis vectors'' have odd number
of nilpotents spread over $SO(13,1)$. They appear in families.


\appendix
\section{Useful table}
\label{usefultable}

This is the copy of Table~1, appearing in Ref.~\cite{n2024NPB}.

In this appendix, the even and odd ``basis vectors''
are presented for the dimension of the internal space $d=(5+1)$, needed 
in particular in Sect.~\ref{feynman}. %


Table~\ref{Table Clifffourplet.}
presents $2^{d=6}=64$ ``eigenvectors" of the Cartan subalgebra,
Eq.~(\ref{cartangrasscliff}), members of the odd and even
``basis vectors'' which are the superposition of odd,
${\hat b}^{m \dagger}_f$, $({\hat b}^{m \dagger}_f)^{\dagger})$,
and even, ${}^{I}{\cal A}^{m}_f$, ${}^{II}{\cal A}^{m}_f$, products
of $\gamma^{a}$'s, needed in Subects.~\ref{fermionbosonstates},
\ref{Feynmanoursandrest}. Table~\ref{Table Clifffourplet.} is presented in
several papers~(\cite{n2023NPB,nh2021RPPNP}, and references therein).


%

\begin{table*}
\begin{small}
\caption{\label{Table Clifffourplet.}  This table, taken from~\cite{n2023NPB}, represents $2^d=64$ ``eigenvectors" of the Cartan subalgebra, Eq.~(\ref{cartangrasscliff}),
members of odd and even ``basis vectors'' which are the superposition of odd 
and even products of $\gamma^{a}$'s  in $d=(5+1)$-dimensional internal space,
divided into four groups. The first group, $odd \,I$, is chosen to represent ``basis vectors", named  ${\hat b}^{m \dagger}_f$,
appearing in $2^{\frac{d}{2}-1}=4$ 
``families" ($f=1,2,3,4$), each ''family'' having $2^{\frac{d}{2}-1}=4$
``family'' members ($m=1,2,3,4$).
The second group, $odd\,II$, contains Hermitian conjugated partners of the first
group for each ``family'' separately, ${\hat b}^{m}_f=$
$({\hat b}^{m \dagger}_f)^{\dagger}$. Either $odd \,I$ or $odd \,II$ are products
of an odd number of nilpotents (one or three) and projectors (two or none).
The ``family" quantum numbers of ${\hat b}^{m \dagger}_f$, that is the eigenvalues of
$(\tilde{S}^{03}, \tilde{S}^{12},\tilde{S}^{56})$, appear for the first {\it odd I }
group above each ``family", the quantum
numbers of the ``family'' members $(S^{03}, S^{12}, S^{56})$ are 
written in the last three columns. 
For the Hermitian conjugated partners of {\it odd I}, presented in the group {\it odd II},
the quantum numbers $(S^{03}, S^{12}, S^{56})$ are presented above each group of the
Hermitian conjugated partners, the last three columns 
tell eigenvalues of $(\tilde{S}^{03}, \tilde{S}^{12},\tilde{S}^{56})$.
Each of the two groups with the even number of $\gamma^a$'s,
{\it even \,I} and {\it even \,II},
 has their Hermitian conjugated partners within its group.
The quantum numbers $f$, that is the eigenvalues of
$(\tilde{S}^{03}, \tilde{S}^{12},\tilde{S}^{56})$, are written above column of
four members, the quantum numbers of the members, $(S^{03}, S^{12}, S^{56})$, are
written in the last three columns. To find the quantum numbers of $({\cal {\bf S}}^{03},
{\cal {\bf S}}^{12}, {\cal {\bf S}}^{56})$ one has to take into account that
${\cal {\bf S}}^{ab}$ $= S^{ab} + \tilde{S}^{ab} $.
 \vspace{2mm}}
 \end{small}
\begin{tiny}
\begin{center}
  \begin{tabular}{|c|c|c|c|c|c|r|r|r|}
\hline
$ $&$$&$ $&$ $&$ $&&$$&$$&$$\\
$''basis\, vectors'' $&$m$&$ f=1$&$ f=2 $&$ f=3 $&
$ f=4 $&$$&$$&$$\\ 
$(\tilde{S}^{03}, \tilde{S}^{12}, \tilde{S}^{56})$&$\rightarrow$&$(\frac{i}{2},- \frac{1}{2},-\frac{1}{2})$&$(-\frac{i}{2},-\frac{1}{2},\frac{1}{2})$&
$(-\frac{i}{2},\frac{1}{2},-\frac{1}{2})$&$(\frac{i}{2},\frac{1}{2},\frac{1}{2})$&$S^{03}$
 &$S^{12}$&$S^{56}$\\ 
\hline
$ $&$$&$ $&$ $&$ $&&$$&$$&$$\\ 
$odd \,I\; {\hat b}^{m \dagger}_f$&$1$& 
$\stackrel{03}{(+i)}\stackrel{12}{[+]}\stackrel{56}{[+]}$&
                        $\stackrel{03}{[+i]}\stackrel{12}{[+]}\stackrel{56}{(+)}$ & 
                        $\stackrel{03}{[+i]}\stackrel{12}{(+)}\stackrel{56}{[+]}$ &  
                        $\stackrel{03}{(+i)}\stackrel{12}{(+)}\stackrel{56}{(+)}$ &
                        $\frac{i}{2}$&$\frac{1}{2}$&$\frac{1}{2}$\\ 
$$&$2$&    $[-i](-)[+] $ & $(-i)(-)(+) $ & $(-i)[-][+] $ & $[-i][-](+) $ &$-\frac{i}{2}$&
$-\frac{1}{2}$&$\frac{1}{2}$\\ 
$$&$3$&    $[-i] [+](-)$ & $(-i)[+][-] $ & $(-i)(+)(-) $ & $[-i](+)[-] $&$-\frac{i}{2}$&
$\frac{1}{2}$&$-\frac{1}{2}$\\ 
$$&$4$&    $(+i)(-)(-)$ & $[+i](-)[-] $ & $[+i][-](-) $ & $(+i)[-][-]$&$\frac{i}{2}$&
$-\frac{1}{2}$&$-\frac{1}{2}$\\ 
\hline
$ $&$$&$ $&$ $&$ $&&$$&$$&$$\\ 
$(S^{03}, S^{12}, S^{56})$&$\rightarrow$&$(-\frac{i}{2}, \frac{1}{2},\frac{1}{2})$&
$(\frac{i}{2},\frac{1}{2},-\frac{1}{2})$&
$(\frac{i}{2},- \frac{1}{2},\frac{1}{2})$&$(-\frac{i}{2},-\frac{1}{2},-\frac{1}{2})$&
$\tilde{S}^{03}$
&$\tilde{S}^{12}$&$\tilde{S}^{56}$\\ 
&&
$\stackrel{03}{\;\,}\;\;\,\stackrel{12}{\;\,}\;\;\,\stackrel{56}{\;\,}$&
$\stackrel{03}{\;\,}\;\;\,\stackrel{12}{\;\,}\;\;\,\stackrel{56}{\;\,}$&
$\stackrel{03}{\;\,}\;\;\,\stackrel{12}{\;\,}\;\;\,\stackrel{56}{\;\,}$&
$\stackrel{03}{\;\,}\;\;\,\stackrel{12}{\;\,}\;\;\,\stackrel{56}{\;\,}$&
&&\\
\hline
$ $&$$&$ $&$ $&$ $&&$$&$$&$$\\ 
$odd\,II\; {\hat b}^{m}_f$&$1$ &$(-i)[+][+]$ & $[+i][+](-)$ & $[+i](-)[+]$ & $(-i)(-)(-)$&
$-\frac{i}{2}$&$-\frac{1}{2}$&$-\frac{1}{2}$\\ 
$$&$2$&$[-i](+)[+]$ & $(+i)(+)(-)$ & $(+i)[-][+]$ & $[-i][-](-)$&
$\frac{i}{2}$&$\frac{1}{2}$&$-\frac{1}{2}$\\ 
$$&$3$&$[-i][+](+)$ & $(+i)[+][-]$ & $(+i)(-)(+)$ & $[-i](-)[-]$&
$\frac{i}{2}$&$-\frac{1}{2}$&$\frac{1}{2}$\\ 
$$&$4$&$(-i)(+)(+)$ & $[+i](+)[-]$ & $[+i][-](+)$ & $(-i)[-][-]$&
$-\frac{i}{2}$&$\frac{1}{2}$&$\frac{1}{2}$\\ 
\hline
&&&&&&&&\\ 
\hline
$ $&$$&$ $&$ $&$ $&&$$&$$&$$\\ 
$(\tilde{S}^{03}, \tilde{S}^{12}, \tilde{S}^{56})$&$\rightarrow$&
$(-\frac{i}{2},\frac{1}{2},\frac{1}{2})$&$(\frac{i}{2},-\frac{1}{2},\frac{1}{2})$&
$(-\frac{i}{2},-\frac{1}{2},-\frac{1}{2})$&$(\frac{i}{2},\frac{1}{2},-\frac{1}{2})$&
$S^{03}$&$S^{12}$&$S^{56}$\\ 
&& 
$\stackrel{03}{\;\,}\;\;\,\stackrel{12}{\;\,}\;\;\,\stackrel{56}{\;\,}$&
$\stackrel{03}{\;\,}\;\;\,\stackrel{12}{\;\,}\;\;\,\stackrel{56}{\;\,}$&

$\stackrel{03}{\;\,}\;\;\,\stackrel{12}{\;\,}\;\;\,\stackrel{56}{\;\,}$&
$\stackrel{03}{\;\,}\;\;\,\stackrel{12}{\;\,}\;\;\,\stackrel{56}{\;\,}$&
&&\\ 
\hline
$ $&$$&$ $&$ $&$ $&&$$&$$&$$\\ 
$even\,I \; {}^{I}{\cal A}^{m}_f$&$1$&$[+i](+)(+) $ & $(+i)[+](+) $ & $[+i][+][+] $ & $(+i)(+)[+] $ &$\frac{i}{2}$&
$\frac{1}{2}$&$\frac{1}{2}$\\ 
$$&$2$&$(-i)[-](+) $ & $[-i](-)(+) $ & $(-i)(-)[+] $ & $[-i][-][+] $ &$-\frac{i}{2}$&
$-\frac{1}{2}$&$\frac{1}{2}$\\ 
$$&$3$&$(-i)(+)[-] $ & $[-i][+][-] $ & $(-i)[+](-) $ & $[-i](+)(-) $&$-\frac{i}{2}$&
$\frac{1}{2}$&$-\frac{1}{2}$\\ 
$$&$4$&$[+i][-][-] $ & $(+i)(-)[-] $ & $[+i](-)(-) $ & $(+i)[-](-) $&$\frac{i}{2}$&
$-\frac{1}{2}$&$-\frac{1}{2}$\\ 
\hline
$ $&$$&$ $&$ $&$ $&&$$&$$&$$\\ 
$(\tilde{S}^{03}, \tilde{S}^{12}, \tilde{S}^{56})$&$\rightarrow$&
$(\frac{i}{2},\frac{1}{2},\frac{1}{2})$&$(-\frac{i}{2},-\frac{1}{2},\frac{1}{2})$&
$(\frac{i}{2},-\frac{1}{2},-\frac{1}{2})$&$(-\frac{i}{2},\frac{1}{2},-\frac{1}{2})$&
$S^{03}$&$S^{12}$&$S^{56}$\\ 
&& 
$\stackrel{03}{\;\,}\;\;\,\stackrel{12}{\;\,}\;\;\,\stackrel{56}{\;\,}$&
$\stackrel{03}{\;\,}\;\;\,\stackrel{12}{\;\,}\;\;\,\stackrel{56}{\;\,}$&
$\stackrel{03}{\;\,}\;\;\,\stackrel{12}{\;\,}\;\;\,\stackrel{56}{\;\,}$&
$\stackrel{03}{\;\,}\;\;\,\stackrel{12}{\;\,}\;\;\,\stackrel{56}{\;\,}$&
&&\\ 
\hline
$ $&$$&$ $&$ $&$ $&&$$&$$&$$\\ 
$even\,II \; {}^{II}{\cal A}^{m}_f$&$1$& $[-i](+)(+) $ & $(-i)[+](+) $ & $[-i][+][+] $ & 
$(-i)(+)[+] $ &$-\frac{i}{2}$&
$\frac{1}{2}$&$\frac{1}{2}$\\ 
$$&$2$&    $(+i)[-](+) $ & $[+i](-)(+) $ & $(+i)(-)[+] $ & $[+i][-][+] $ &$\frac{i}{2}$&
$-\frac{1}{2}$&$\frac{1}{2}$ \\ 
$$&$3$&    $(+i)(+)[-] $ & $[+i][+][-] $ & $(+i)[+](-) $ & $[+i](+)(-) $&$\frac{i}{2}$&
$\frac{1}{2}$&$-\frac{1}{2}$\\ 
$$&$4$&    $[-i][-][-] $ & $(-i)(-)[-] $ & $[-i](-)(-) $ & $(-i)[-](-) $&$-\frac{i}{2}$&
$-\frac{1}{2}$&$-\frac{1}{2}$\\ 
\hline
 \end{tabular}
\end{center}
\end{tiny}
\end{table*}
%



\begin{thebibliography}{}
\bibitem{n2024NPB}
   N. S. Manko\v c Bor\v stnik,
 "Can the ``basis vectors'', describing the internal spaces of fermion and boson fields with the Clifford odd 
(for fermion) and Clifford even (for boson) objects, explain interactions among fields, with gravitons 
included?" 
[arxiv: 2407.09482].
\end{thebibliography}
\end{document}